\documentclass{article}
\usepackage{geometry}

\usepackage{amsmath,amssymb}

\usepackage{changepage}

\usepackage[utf8x]{inputenc}

\usepackage{textcomp,marvosym}

\usepackage{cite}

\usepackage{nameref,hyperref}

\usepackage{microtype}
\DisableLigatures[f]{encoding = *, family = * }

\usepackage[table]{xcolor}

\usepackage{array}

\usepackage{lipsum}
\usepackage{multirow}
\newcolumntype{L}[1]{>{\raggedright\let\newline\\\arraybackslash\hspace{0pt}}m{#1}}
\newcolumntype{C}[1]{>{\centering\let\newline\\\arraybackslash\hspace{0pt}}m{#1}}
\newcolumntype{R}[1]{>{\raggedleft\let\newline\\\arraybackslash\hspace{0pt}}m{#1}}
\usepackage{text comp}
\usepackage{color}

\newcolumntype{+}{!{\vrule width 2pt}}

\newlength\savedwidth

\usepackage[aboveskip=1pt,labelfont=bf,labelsep=period,justification=raggedright,singlelinecheck=off]{caption}

\makeatletter
\renewcommand{\@biblabel}[1]{\quad#1.}
\makeatother

\date{}

\usepackage{graphicx}

\usepackage{authblk}
\usepackage{blindtext}
\usepackage{ragged2e}

\begin{document}

\title{Meaningless comparisons lead to false optimism in medical machine learning}

\author[1]{\rm Orianna DeMasi}
\author[2, 3]{\rm Konrad Kording}
\author[1]{\rm Benjamin Recht}
\affil[1]{Department of Electrical Engineering and Computer Sciences, University of California Berkeley, Berkeley, CA, USA}
\affil[2]{Department of Bioengineering, University of Pennsylvania, Philadelphia, PA, USA}
\affil[3]{Department of Neuroscience, University of Pennsylvania, Philadelphia, PA, USA}
\affil[ ]{\textit {odemasi@berkeley.edu, kording@upenn.edu, brecht@berkeley.edu}}
\date{\today} 

\maketitle

\section*{Abstract}
A new trend in medicine is the use of algorithms to analyze big datasets, e.g. using everything your phone measures about you for diagnostics or monitoring. However, these algorithms are commonly compared against weak baselines, which may contribute to excessive optimism. To assess how well an algorithm works, scientists typically ask how well its output correlates with medically assigned scores. Here we perform a meta-analysis to quantify how the literature evaluates their algorithms for monitoring mental wellbeing. We find that the bulk of the literature ($\sim$77\%) uses meaningless comparisons that ignore patient baseline state. For example, having an algorithm that uses phone data to diagnose mood disorders would be useful. However, it is possible to over 80\% of the variance of some mood measures in the population by simply guessing that each patient has their own average mood - the patient-specific baseline. Thus, an algorithm that just predicts that our mood is like it usually is can explain the majority of variance, but is, obviously, entirely useless. Comparing to the wrong (population) baseline has a massive effect on the perceived quality of algorithms and produces baseless optimism in the field. To solve this problem we propose ``user lift'' that reduces these systematic errors in the evaluation of personalized medical monitoring.


\section*{Introduction}

Health care should be tailored to individuals to maximize their wellbeing and health~\cite{collins15}. There is considerable hope that data collected from emerging data sources, such as smartphones and smartwatches, can be used to extract medical information and thus improve the tailoring of monitoring, diagnostics, and treatments for personalizing health care~\cite{darcy16}. In particular, mental health care could particularly benefit from automated monitoring, as many mental health conditions need long-term monitoring and clinical monitoring is expensive, but automatically tracking a user with ubiquitous sensors is cheap~\cite{calvo16, marzano15, olff15}.

Machine learning algorithms are commonly being used in an attempt to extract medical information from easy to collect data sources~\cite{benzeev15, canzian15, Pantelopoulos10, saeb15}. These algorithms are attractive as, by automating information extraction, they promise to provide rich analyses cheaply and objectively based on collected data. Machine learning works by taking data that are easy to collect, building a model, and then using the model to make predictions for data that are harder to collect~\cite{friedman01}. As an example, social media posts may be used to predict individuals' depressive symptoms or future suicidal ideation~\cite{dechoudhury14, dechoudhury16}. However, without sufficient evaluation, the outputs of algorithms may be meaningless and mislead clinicians.

Whenever algorithms are used to make predictions, they must carefully be evaluated to ensure that their predictions meaningfully represent medically relevant information. Evaluation must be specified for each problem~\cite{witten16}. For example, if an algorithm is being used to predict one of two things, such as whether a patient is depressed, then it could be evaluated by the percent of predictions that are correct~\cite{saeb15}. Alternatively, it could be evaluated by the percent of times that it correctly identified depression, i.e. sensitivity or true positive rate, and ascribe less importance to false positives~\cite{boyko94}. In contrast, if an algorithm is trying to predict a value, such as someone's level of depressive symptoms, one could consider the degree to which predictions differ, i.e., the mean squared error. There are myriad additional methods for evaluating algorithms because, without sufficiently evaluating algorithms, it is easy to generate misplaced optimism about the utility of algorithms~\cite{fawcett06, he09, powers11, weiss04}.

Regardless of how the correctness of an algorithm is quantified, algorithms must be compared to a baseline approach that simply makes guesses to prove that the algorithm makes better predictions than guessing. For example, if an algorithm is trying to predict a rare event, such as a mental breakdown or suicide, an approach that simply guesses that the event never happens will usually be correct and thus will have high accuracy~\cite{kessler16, murphy84, tran13}. However, such an approach is entirely useless for medicine. If algorithms are not compared with reasonable guesses, the accuracy of the algorithm's predictions can appear to be good, when in reality the algorithm is doing no better than guessing and is thus medically useless.

Here we review, for modeling of longitudinal individual state, what baselines algorithms are commonly compared against and how much of the apparent success of algorithms can be ascribed to poor comparisons. We focus on the example of mental wellbeing and demonstrate in two popular datasets that individuals exhibit little variance over time. Typical wellbeing prediction algorithms seem to work well, but we find that this is simply because they are basically always guessing individuals' personal average states. This example highlights how  falsely optimistic results can easily be obtained by comparing machine learning with population as opposed to personal baselines. We perform a systematic literature review and find that most studies ($\sim$77\%) compare with the population baseline. By not comparing with personal baselines, studies are prone to making falsely optimistic conclusions that can unintentionally mislead researchers' perspectives and delay progress on important medical applications. We argue for a new measure, ``user lift," that measures the benefit of an algorithm relative to the single-person model.

\section*{Methods}

\subsection*{Algorithm Evaluation}

There are many ways to evaluate how good an algorithm's predictions are~\cite{he09, friedman01, weiss04, witten16}. The general approach is a two step process of measuring an algorithm's error, or how inaccurate its predictions are, and then comparing the algorithm's error with the error of simply guessing answers. These guesses form a baseline approach and could be specific to each patient, or they could use other trivial factors, e.g. the time of the day. Regardless, because it is totally useless for medicine to simply guess answers based on subject and other trivial factors, algorithms must have lower error than such baselines to be of any use.

\subsection*{Algorithm Error}

To evaluate how well an algorithm predicts a binary outcome, e.g., whether an individual is having a happy vs. sad or stressed vs. relaxed day, we consider the classical measure of prediction error. Prediction error is the percent of observations that were incorrectly predicted (percent incorrect). To evaluate how well an algorithm predicts an individual's level of happiness or stress, we consider the root mean squared error (RMSE), which considers how different the predicted levels are from the true reported levels~\cite{friedman01}. With both prediction error and RMSE, lower values indicate that an algorithm is doing better at predicting an individual's state. Higher values indicate more significant prediction error.

\subsection*{Baselines}

We consider two baseline methods that simply guess how an individual is doing: personal baselines and population baselines. Both baselines are simple approaches that always guess individuals are at the same state. The personal baseline always guesses that each individual is at a constant state, but that state can differ between individuals. The population baseline predicts that all individuals are always at the same state. 

When an algorithm is attempting to predict whether an individual is having a stressed (or happy) day or not, we consider personal baseline error, which is the prediction error of always guessing that each individual is always at their most frequently reported state (mode). We also consider the population baseline error, which is the error of always guessing that all individuals are always at the most frequently reported state of the population (mode).

When an algorithm is attempting to predict an individual's level of happiness or stress, we consider the personal baseline RMSE, which is how far predicted levels were from always guessing each individual to be at their average level of stress or happiness. RMSE indicates a model with higher error and thus worse predictions. We also consider the population baseline RMSE, which is the RMSE of always guessing that all individuals are always at the average state of the population.

\subsection*{User lift}

We propose the measure of user lift as a way to evaluate whether an algorithm is making better predictions than simply guessing an individual's state. The user lift is the improvement of an algorithm's predictions over the personal baseline, or the amount that error is decreased by adding better features and a model. User lift is the difference between personal baseline error and model error in RMSE or in prediction error (personal baseline error - model error). The user lift can be thought of as the increase in accuracy of an algorithm over the null accuracy of guessing an individual to be at their average state. The average user lift is the mean user lift across the individuals in the dataset.

\subsection*{User lift framework}
As a stricter measure of whether algorithms have any utility, we suggest the user lift framework instead of comparison with a single weak baseline, such as the population baseline. With this framework, researchers calculate user lift for each study participant. The user lift quantifies whether an algorithm is better than the simple personal baseline on each user. We propose then reporting descriptive statistics on the distribution of user lift and utilizing statistical tests to determine whether the average user lift is greater than zero. Nonparametric permutation tests are appropriate and powerful tests for considering whether a single sample, such as of user lifts on study participants, has a mean greater than zero. A permutation test is appropriate here so that no assumptions on distributions are needed. While other nonparametric tests, such as the paired Wilcoxon signed-rank test, may be appropriate for comparing two samples, permutation tests have been reported to be more reliable than paired non-parametric tests \cite{kempthorne69, smucker07}. 

\subsection*{Machine Learning Example: Predicting Subjective State from Location and Mobility}

We present an example of how falsely optimistic conclusions can be reached about algorithms' performance. For this example, we follow previous works that have used and suggested that smartphone GPS location data can predict individuals' mental wellbeing~\cite{canzian15, jaques15, saeb15}.

\subsection*{Datasets}

We consider two well established datasets that are freely available. Both datasets collected individuals' smartphone data, specifically GPS location, and their stress and happiness levels. The StudentLife dataset~\cite{wang14} followed a cohort of students at an American university during the course of a semester. Data that was collected included daily measures of stress on a five point Likert scale. Of the initial 48 students with data accessible, we consider data for the 15 students who had sufficient data available: stress level and at least 35 GPS location observations for at least 30 days of the study period.

The second dataset we consider, the MIT Friends and Family dataset~\cite{aharony11}, resulted from a project that collected various types of data on a cohort of university affiliates and their families at another American university. The data collected included daily wellbeing measures. Here we consider the nine point Likert scale of happiness and seven point Likert scale of stress that were collected. Of the 116 participants included in the available dataset we consider data for the 31 individuals who had measurements of stress or happiness, respectively, for at least 30 study days and at least 35 GPS location measurements on those days.

\subsection*{Data Processing: Location and Mobility Features}

To derive meaningful features of location and mobility, we follow three previous studies~\cite{canzian15, jaques15, saeb15}. All features from these studies that were reproducible (due to the data available) were included. Before constructing features, we used two preprocessing methods. 

The first preprocessing method fit a Gaussian Mixture Model (GMM) to all of the location samples for each participant collected to identify locations frequented by participants~\cite{jaques15}. The number of clusters was chosen to be the number, up to twenty maximum, that minimized the Bayesian Information Criterion~\cite{schwarz78}. It was assumed that participants would frequent at most twenty locations during the course of the study. The home location of a participant was determined to be the location where the participant spent the majority of their time during the evening hours (11pm - 6am) and the work location was similarly determined to be where the participant spent the majority of their time during working hours (11am - 4pm). In contrast to prior work, we did not interpolate the location observations to a regular time sampling, as we did not find this beneficial to prediction accuracy~\cite{jaques15}. We consider this first set of clusters to be the full clustering.

The second preprocessing method used K-means clustering on stationary points only~\cite{arthur07, saeb15}. The StudentLife dataset included a prediction of whether the participant was moving or stationary at each observation, but the Friend and Family dataset did not. To determine whether participants in the Friends and Family dataset were stationary at each observation, we approximated movement speed with the time derivative at each observation and used a threshold. We attempted to set the threshold to be about 1km/h~\cite{saeb15}. We consider this second set of clusters as the stationary clustering and the night cluster to be the cluster where each individual spent the most time between midnight and 6am. 

To protect participants' anonymity, the GPS location data in the Friends and Family dataset was subjected to an affine transform before being released. Because this transform purposefully changes the space, but collinearity should be preserved, we approximated features in one dimension on the Friends and Family dataset.

Utilizing the two set of location clusters that resulted from the GMMs and Kmeans, the features of mobility and location that we derived for each participant each day of the study are as follows:

\begin{enumerate}

\item{ The fraction of a day that a participant spent not stationary.}
\item{ The average displacement of a participant between two observations during the day, i.e., average speed. }
\item{ The standard deviation of displacements between points.}
\item{ The location variance (on log scale), i.e., the sum of the variance of location coordinates in each dimension.}
\item{ The ``circadian movement" of a participant~\cite{saeb15}, which we adapted to our daily monitoring setting as the Euclidean distance of the vector of fraction of time a participant spent in each of their stationary location clusters with the participant's mean location distribution. The mean location distribution of a participant was calculated as the average fraction of a day that a participant would spend in each stationary location cluster during the study. }
\item{ The location entropy, which was calculated as the entropy of the vector where each entry represented the fraction of the day that a participant spent in each stationary location cluster. }
\item{ The radius of minimum circle enclosing the participant's location samples. }
\item{ The fraction of time a participant spent at their GMM home cluster. }
\item{ The fraction of time a participant spent at their GMM work cluster. }
\item{ The fraction of time a participant spent at their stationary night cluster. }
\item{ The log likelihood of a day from the GMM to estimate how routine the day was.}
\item{ The AIC and BIC of the GMM evaluated with the day's coordinates, to also determine how typical the day was.}
\item{ The number of GMM clusters visited in a day. }
\item{ The number of stationary clusters visited in a day.}
\end{enumerate}

\subsection*{Experimental Framework}

We present two prediction tasks: 
\begin{enumerate}
\item{ Predicting whether a participant was happy or stressed or not on a given day.}
\item{ Predicting the average level of happiness or stress that a participant reported on a given day.}
\end{enumerate}

To construct levels of stress or happiness on a given day, we average all the Likert scale responses that a participant reported on that day.  Whether the participant was happy (or stressed) or not is defined by a threshold on the daily average on a value to distinguish when students reported any stress versus no stress. For the StudentLife user inputs, we use  ``A little stressed" as the threshold. For the Friends and Family dataset, we use the middle value of the Likert scale as the threshold, as the Friends and Family scales were defined from negative to positive values, where the middle value was supposed to indicate a neutral state.

For both problems, we attempt to predict the stress or happiness from the location and mobility data with a variety of standard machine learning methods. For regression we consider: linear regression with an Elastic Net penalty, and Lasso regression~\cite{tibshirani96, zou05}. For the binary classification task, we consider: logistic regression with L2 penalty, support vector machines with radial basis function kernels, and random forests~\cite{breiman01, friedman01}. Hyperparameters were chosen with 10-fold cross-validation on the training data. The methods that return the lowest error are presented. 

We consider both population models, which could also be referred to as global, general, or all-user models and utilize all the individuals' data to make predictions, and personal models, which use only a single individual's data to make predictions for that individual. Prediction error is measured with leave-one-out cross-validation, which is commonly used for estimating an algorithm's prediction error~\cite{friedman01}. 
To perform leave-one-out cross-validation on population models, we combine data from all of the participants into a single set. Then one observation is withheld, a model is trained on all of the other observations, and then that model is used to make a prediction for the held out observation. The process is repeated until every observation has been withheld exactly once. The model error reported from this process is the average error across the predictions for each data point. Population models assume that some of each participant's data is seen during training, in addition to data from other participants~\cite{saeb16}. For personal models, we similarly hold out one observation, but we only train on the remaining observations of that individual's data and then repeat only for the number of observations that we have on that individual. Personal models only attempt to extrapolate predictions for an individual from their own data. Alternative cross validation schemes, such as N-fold, offered no benefit to the results, so are omitted for brevity.

\subsection*{Literature Review}

In addition to an example on two real datasets of how false machine learning results can be arrived at by comparing to weak baseline models, we perform a systematic literature review to investigate how algorithms are commonly evaluated and whether baselines are sufficiently reported. Our literature review took three steps:

\begin{enumerate}
\item{ Find relevant literature. }
\item{ Establish whether a baseline (personal or population) was compared with.}
\item{ Identify the error of the baselines and the best reported machine learning algorithms.}
\end{enumerate}

\subsection*{Finding Relevant Literature}
While baselines are needed to evaluate all machine learning algorithms on personal data, we make our literature review tractable by focusing on studies similar to the machine learning example we present. We utilize GoogleScholar to find publications that attempted to automatically infer an individual's subjective states, similar to the example we presented. The studies we include meet the following criteria:

\begin{itemize}
\item Relate to subjective personal data, as denoted by having one of the following words in the title: depression, depressive, stress, mood, mental, happiness, or wellness. 
\item Attempt a machine learning prediction task and report prediction accuracy by having the following word ``accuracy" somewhere in the text of the publication.
\item Attempt prediction on participants' longitudinal data, where personal baselines are defined, by containing the words ``participant" or ``user". 
\item Collect data from sensors by requiring one of the following words to be included somewhere in the text: smartphones, sensor, sensors, or sensing.
\item Were published since 2010. 
\end{itemize}

Because of a particularly strong focus on stress in previous work, we break the query into two queries: one that requires the word ``stress" to be in the title and another search that requires any of the other wellbeing words to be in the title. We perform this joint search with the following GoogleScholar queries: 
\newline

\textbf{\textit{(participant OR user) accuracy (sensor OR sensors OR smartphones OR sensing)intitle:stress (from 2010)}}
\newline 

\textbf{\textit{(participant OR user) accuracy (sensor OR sensors OR smartphones OR sensing) (intitle:mental OR intitle:depression OR intitle:depressive OR intitle:mood OR intitle:happiness) (from 2010)}}
\newline

To be considered relevant, studies need to attempt to predict user input data of users' subjective state from other collected data, i.e. sensors. Examples of studies that are returned by our query, but are excluded from our analysis are: 

\begin{itemize}
\item Correlational analyses that reveal certain data or behaviors are correlated with subjective state.
\item Studies of one-time user surveys (in contrast to repeated prompts) or where the goal is to separate subjects, i.e., each subject was a data point.
\item Literature summaries or reviews.
\item Randomized control studies of intervention efficacy. 
\item Other evaluations of treatments on subjective state. 
\item Collection and presentation of a dataset collected without a prediction task. 
\item Measurements of behaviors without attempting prediction. 
\item Descriptions of tools and systems implemented with user reviews of the systems. 
\item Non-peer reviewed publications, such as reports and book chapters. 
\item Prediction of non-subjective states, e.g., prediction of labels coded by researchers who intuit what state the user was in from observational data, or labels of stimulus exposure when studies attempted to induce a given emotional state such as stress. 
\end{itemize}

We only consider studies where labels are for multiple observations of a participant's subjective input state. 

\subsection*{Establishing Comparison with a Baseline}
Some studies do not report any baseline model for comparison, so we begin by noting which studies reported a baseline model. For studies that provided sufficient detail, we did the following: 

\begin{itemize}
\item When baseline models are reported we recorded the baseline performance metrics directly from the text and the type of baseline used, e.g., a population baseline or a user baseline.  
\item When baseline models are not provided, but confusion matrices are provided we manually calculate the baseline performance.
\item When individuals baselines are reported, we take the average user baseline performance. 
\item When only mean squared error are reported, we note whether the mean squared error is also provided for a constant baseline. 
\end{itemize}

\subsection*{Comparing the accuracy of different models}
There are a wide variety of performance metrics authors report when evaluating their models. We extract model prediction error for multi-class classification problems according the to following criteria:

\begin{itemize}
\item When results are broken down for personal models by individual, the average is used. 
\item When accuracy results are given for multiple objectives, e.g., different dimensions of mood, the best results for each objective is recorded. 
\item When multiple feature sets and models were tried, only the best performing model is considered. Models that utilize user input as features were excluded when possible. 
\item The number of folds in the cross validation scheme used is not incorporated into our analysis. We considered 10-fold, leave-one-out, and leave-user-out cross validation schemes to all be ``population" models. Both 10-fold and leave-one-out cross validations on personal data only are considered to be ``personal" models.
\item The uniform baseline was calculated by noting the number of classes that the study reported using in their measurement scale. 
\end{itemize}

\section*{Results}

We want to consider to what extent  the choice of baselines matter in medical machine learning and how baselines are used in practice. To quantify the importance of baselines, we use two publicly available datasets and compare the performance of machine learning algorithms to two different baselines: a population baseline and a personal baseline. More specifically, we use the StudentLife~\cite{wang14} and Friends and Family~\cite{aharony11} datasets and analyze machine learning predictions of stress and happiness, which we compare to both personal and population baselines. To understand how the field generally uses baselines, we perform a systematic literature review. These two complementary analyses will allow us to meaningfully inform the debate about machine learning in medicine.

\begin{figure}[!h]
\begin{center}
\includegraphics[width=\linewidth]{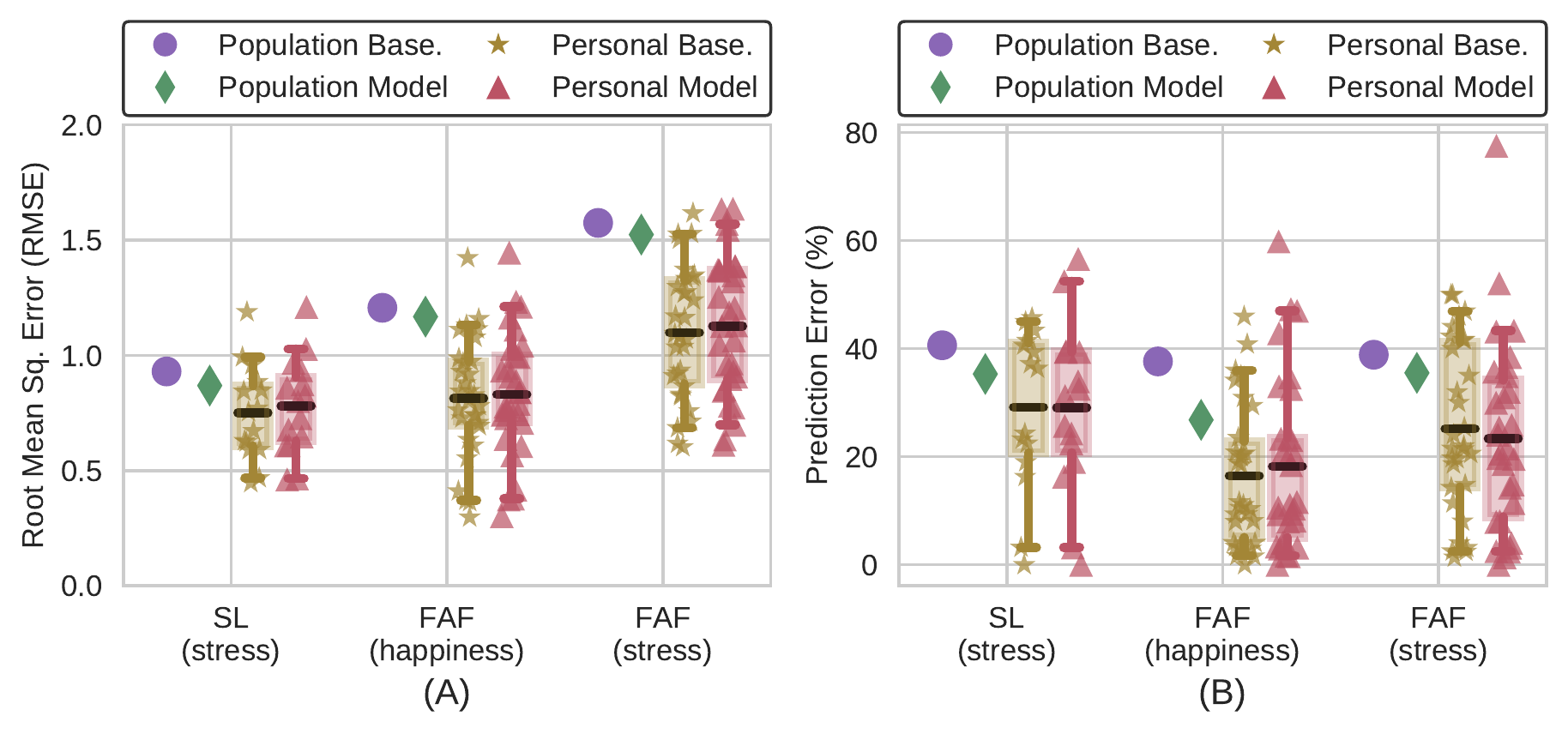}
\caption{{\bf Results of machine learning models on StudentLife (SL) and Friends and Family (FAF) datasets.} 
Bars represent the 5th and 95th percentiles, black lines indicate means, and boxes indicate the 1st and 3rd quartiles of error incurred on individuals. Personal models yield lower error than population models and population baselines, which often leads researchers to the conclusion that personal models are successful. Comparing personal models with personal baselines reveals that their error is no lower, so algorithms are doing no better than predicting individuals to be their most frequently reported state. The models presented are those with lowest error.  }
\label{fig1}
\end{center}
\end{figure}

Initially, we find, that individual subjects have little variance over time, relative to the variance across the population, i.e. low personal baseline error relative to higher population baseline error (Fig~\ref{fig1}). 
Thus, comparing learned models with population baselines can obscure whether a model is better (lower error) on individuals than constant personal baseline models.
We find the same pattern when we ask about RMSE and binary predictions. This gives us an intuition that guessing each subject's mean value should produce relatively low errors.

Motivated by the intuition that there is little within-subject and more across-subject variance, we now ask how machine learning algorithms compare to the two baselines. In line with prior literature~\cite{canzian15, jaques15}, the algorithms predict whether an individual was having a particularly stressful or happy day from their GPS location and mobility data (Fig~\ref{fig1}). 
Our binary results are comparable (max difference=6\%) to past studies predicting binary stress or emotion from similar datasets~\cite{bogomolov14, bogomolov14_pervasive, carroll13, jaques15, jaques15_nips,  lu12, moturu11}, as are the errors of the personal models~\cite{canzian15, deng12, hernandez11, lu12, sandulescu15, valenza13, valenza14, wu15}. Similarly, the difference between RMSE of personal models and personal baselines is comparable to the differences reported in prior publications~\cite{asselbergs16, likamwa13}. Our algorithms are much better than the population baseline and population models. They are not, however, lower than the personal baselines. This shows how good performance relative to the population baseline can be entirely meaningless.

To prevent comparing with the wrong baseline and to control against obscuring the range of how well algorithms do on individuals with aggregate statistics, we propose using statistical tests with the metric of user lift to prove that an algorithm is doing significantly better than the personal baseline. User lift is the difference of the personal model with the personal baseline, as described above. Positive user lift indicates that a model is better than the personal baseline, that the algorithm's predictions are more accurate than always assuming an individual is at their average state. Indeed, user lift shows that our naive model is useless while our moderately careful model at least adds something (Table~\ref{table1}). 
Using the wrong baseline, may make bad machine learning with a performance that is by any meaningful definition useless seem impressive underscoring the importance of meaningful baselines.

\begin{table}[!b]
\centering
\caption{{\bf Statistical significance for user lift of personal models in Fig~\ref{fig1}.} 
The user lifts are the differences of personal baselines with personal models, in terms of prediction error or RMSE. The p-values are for permutation tests considering whether the user lifts were larger than zero. In every case the user lifts are not significantly greater than zero - the models are not doing better than constant personal baselines.}
\begin{tabular}{l l c |C{2.4cm} C{2.4cm} | C{2cm} c}
\hline
Dataset & Problem & Model & Avg. Personal Baseline Error & Avg. Personal Model Error  & Avg. User Lift (Error) & p-value\\
\hline
\hline
SL - Stress		&binary	&Log.Reg.&	29.19\%&	29.09\%& \textbf{0.10}	& .481\\
FaF - Happiness	&binary	&SVM(rbf)	&	16.51\%&	18.67\%&\textbf{-2.17}	& .967\\
FaF - Stress		&binary	&SVM(rbf)&	25.17\%&	23.35\%& \textbf{1.82}	& .240\\
\hline
SL - Stress		&regression	&Elastic Net	&	0.75&	0.78& \textbf{-0.03}	& .988\\
FaF - Happiness	&regression	&Elastic Net	&	0.81&	0.83&\textbf{-0.02}	& .999\\
FaF - Stress		&regression	&Elastic Net&	1.10&	1.13& \textbf{-0.03}	& 1.000\\
&&&&& \\
\end{tabular}
\label{table1}
\end{table}

\begin{figure}[!t]
\begin{center}
\includegraphics[width=.7\linewidth]{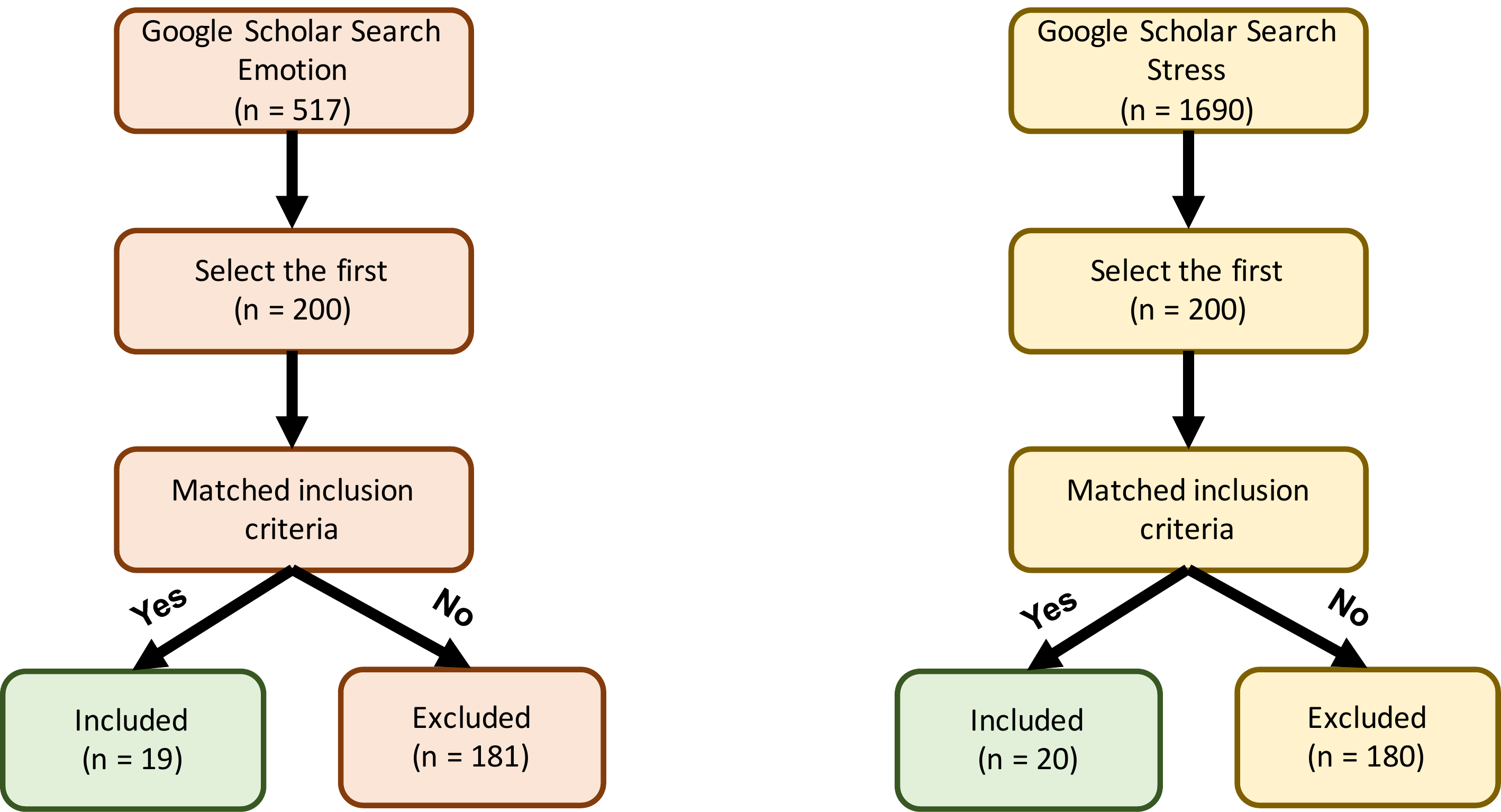}
\caption{{\bf Diagram of literature review process.}}
\label{fig2}
\end{center}
\end{figure}

\begin{figure}[!b]
\begin{center}
\includegraphics[width=\linewidth]{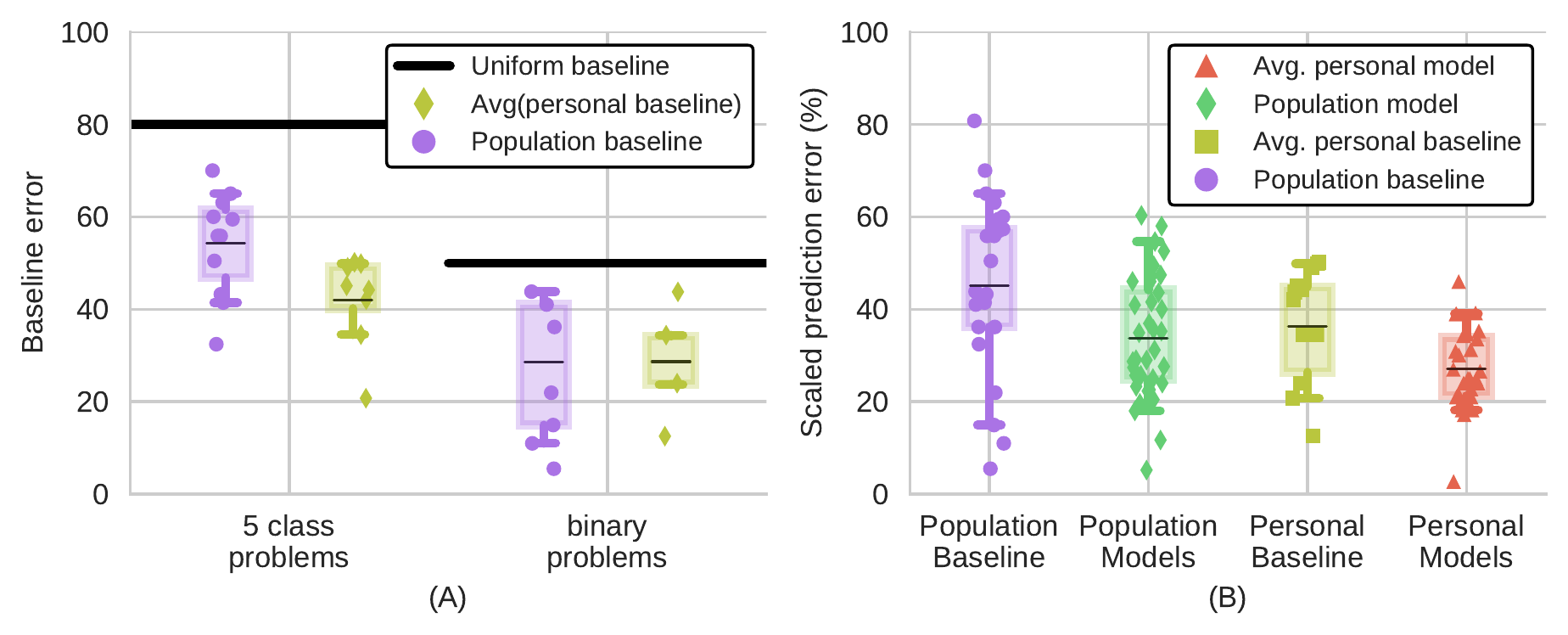}
\caption{{\bf Participant variability and model performance reported in related studies.} Reported results reveal little participant variability and that models do not dramatically improve upon personal baselines. (A) Population and personal baselines reported by studies that had participants report their state on two point and five point scales. The black bars indicate the what the baseline would have been if participants were to report every state equally often, e.g., happy half the time. (B) Population and personal baselines and model error reported in literature reviewed. Performance (prediction error) is scaled by the minimum class imbalance to compare studies that asked participants to report their states on scales with different numbers of points. In both figures boxes denote 1st and 3rd quantiles, bars indicate 5th and 95th percentiles, and lines the average of the markers.}
\label{fig3}
\end{center}
\end{figure}

To understand how algorithms are typically evaluated, we perform a systematic literature review of related studies that attempted to predict emotion and stress from sensed data, such as from smartphones or smartwatches (Fig~\ref{fig2}). 
Just like in our example datasets, participants report surprisingly little variation (Fig~\ref{fig3}A). 
As a result, guessing that an individual was at the same state incurred low personal baseline error and machine learning algorithms typically had only slightly lower error than the personal baseline (Fig~\ref{fig3}B).

\begin{figure}[!h]
\begin{center}
\includegraphics[width=\linewidth]{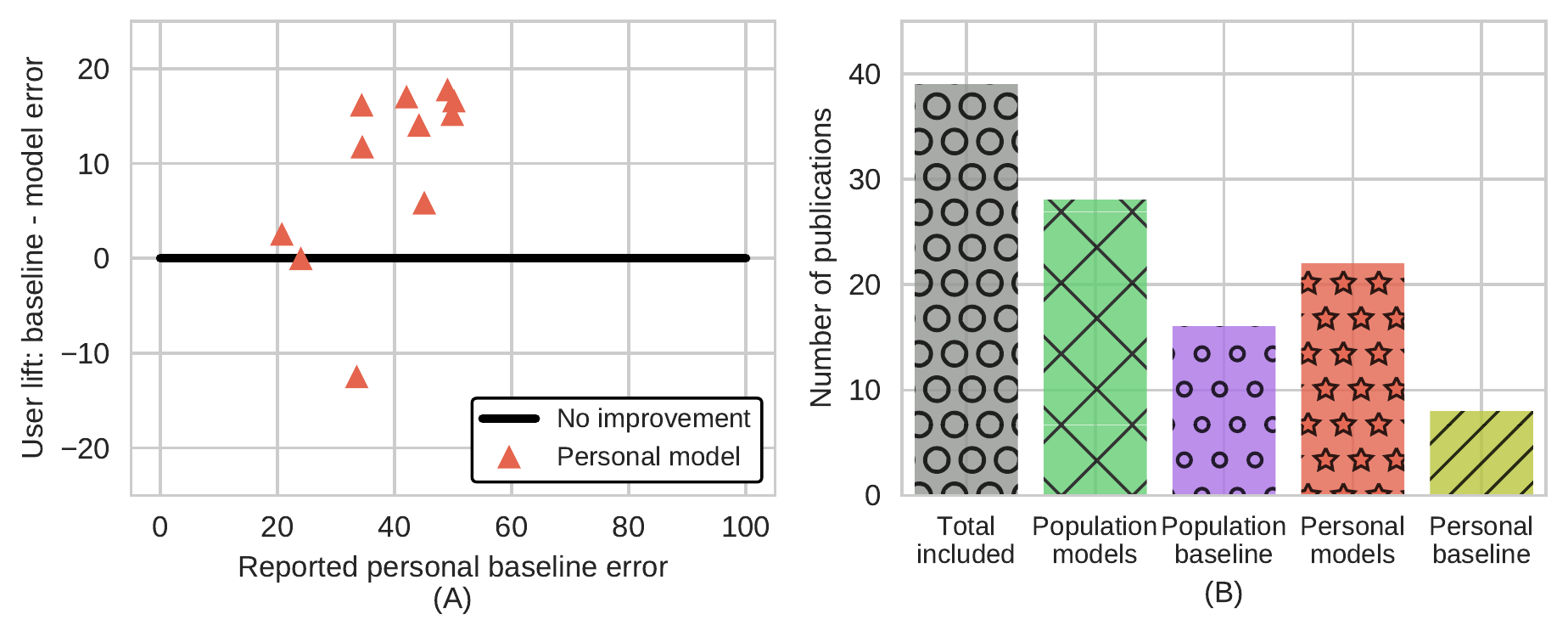}
\caption{{\bf Calculated user lift and prevalence of baselines reported in studies reviewed.} 
(A) User lift calculated for studies where error of baselines and algorithms were both reported. Algorithms sometimes have no improvements over baseline guessing, and these figures are biased to studies that reported sufficient information. (B) Population baselines are reported in roughly half the publications reviewed while personal baselines are infrequently reported (approximately 23\% of publications). }
\label{fig4}
\end{center}
\end{figure}

Studies that do report personal averages sometimes have negative user lift (Fig~\ref{fig4}A).
When personal baselines are reported, they are usually reported in aggregate, which can be misleading by obscuring negative user lift on some individuals. Aggregation also precludes statistical tests on user lift and the only study that did report a statistical rank test on improvement across individuals found that there algorithms were no better than a naive model (using an historical averages of individuals' states)~\cite{asselbergs16}. However, the bulk of studies only report population baselines making it impossible to know if they have any user lift (Fig~\ref{fig4}B). 
 As such, it seems that the bulk of papers have questionable results, at best.

\section*{Discussion}

We have shown, with examples of stress and happiness on two popular datasets how easily machine learning algorithms can appear promising when compared with meaningless baselines. Individuals report surprisingly little variation in state, so always guessing that an individual is at their most frequently reported state is correct most of the time. As a result, when an algorithm is compared with a population baseline that always predicts all users are always at the same state, the algorithm's predictions can seem accurate even if they are no better than predicting each individual to be at their most frequently reported state. Despite the possibility for falsely optimistic results, we found in a systematic literature review that population baselines are commonly compared with in roughly 77\% of publications reviewed. We also find that when personal baselines are reported that the algorithms often add little or nothing over these baselines (and in fact they sometimes do worse). 

A limitation of the datasets that we explored, and most of the literature we reviewed, was that the study cohorts were not clinical populations, the sample size was small, and the study duration was limited. However, the study characteristics of the datasets presented are characteristic of many studies. While target populations for the monitoring we have discussed are typically individuals with mood disorders, study cohorts are frequently small in size and from the general population. It is possible that individuals with mood disorders would report more variability in state than the general public. More variability would reduce the likelihood of falsely optimistic results, but our proposed evaluation method would still be appropriate for showing that algorithms are an improvement over always predicting that individuals are at their average state. Finally, the user lift evaluation framework that we suggest would complement a larger dataset, despite being demonstrated on fewer subjects here.

While we reviewed a representative portion of relevant literature, we had to focus the scope and present a reproducible search that aligned with public datasets. We constructed general search queries to include pertinent studies, but inconsistencies in terminology between communities made it impossible to included all relevant studies and some known related works were not covered. 

In addition to coverage, there were a variety of features that we could not control in the literature review. Studies recruited from disparate populations and had different study protocols. In addition to collecting different data and conducting different analyses, studies reported results in an variety of ways. We did our best to standardize across studies and present results favorably and comparably.   

The proposed user lift evaluation framework is more generally applicable to predicting longitudinal patient state than we have shown here. We have focused our review on a narrow, important application of mental wellbeing, as this is a nascent and exciting application for machine learning algorithms. However, user lift would apply to any application predicting longitudinal data, such as monitoring blood sugar level, body weight, or daily sleep duration. The importance of statistical tests on user lift becomes greater for applications where individuals are expected to exhibit less variation and descriptive statistics must also be reported to quantify the size of any statistically significant user lift. 

While we have calculated personal baselines here over the entire dataset, in principle this is not necessary. Because personal baselines are calculated with respect to individuals' most common state, personal baselines are easy to quickly approximate, with minimal sampling. Personal baselines could potentially vary over an extended period of study, but such scales are outside the scope of most studies and require further investigation.

An ability to predict meaningful personal signals for medical monitoring, such as mental wellbeing, could greatly improve personalized medicine by enabling novel approaches to just in time and personalized interventions. However, we have highlighted some pitfalls of evaluating algorithms for this application that can easily result in falsely optimistic results and unintentionally provide baseless optimism. To reduce falsely optimistic results, we have suggested an alternative evaluation framework using statistical tests on our proposed metric of user lift, which takes an individual-centric approach. As was shown, there is a range of model predictive capability across individuals, so we suggest statistically testing for significant improvement on the population. This framework of evaluation can help researchers to focus efforts and thus help advance progress on this application.


\end{document}